\documentstyle[12pt]{article}
\begin{document}
{\centerline{\bf{\large{Superconducting cosmic string in Brans-Dicke theory}}}}
\vspace{5mm}
{\centerline{\bf{A.A.Sen{\footnote{e-mail: anjan@juphys.ernet.in}}}}
\vspace{2mm}
{\centerline{Relativity and Cosmology Research Center}}
{\centerline{Department of Physics, Jadavpur University}}
{\centerline{Calcutta 700032, India}}
\vspace{5mm}
{\centerline{Abstract}}
\vspace{2mm}
In the present work, the gravitational field of superconducting cosmic string
has been investigated in the context of Brans-Dicke theory of gravity. We have 
presented two kind of solutions for the spacetime in the far field zone of the 
string. When the B-D scalar field is switched off, one of the solutions reduces
to the solution earlier obtained by Moss and Poletti in General Relativity.
\newpage
\vspace{2mm}
Spontaneous symmetry breaking in gauge theories may give rise to some 
topologically trapped regions of a false vacuum namely domain walls, cosmic 
strings or monopoles depending on the dimension of the region\cite{R1}.
Amongst these, cosmic strings have gained a lot of attention in recent years
as a possible seed for galaxy formation. Strings arising from the breaking of
a U(1) local symmetry are called local strings whereas those arising from the
breaking of U(1) global symmetry are called the global strings. The 
gravitational fields of both local and global strings have been investigated
by many authors in recent past \cite{R2}-\cite{R8}.

It has been demonstrated by Witten \cite{R9} that under certain conditions of
local gauge symmtery breaking, cosmic strings might behave as superconductors
whose motion through astronomical magnetic fields can produce interesting 
effects. Peter\cite{R10} has emphasized that superconductivity is a rather
generic feature of cosmic string models. Massive superconducting strings 
may have important role to play in the generation of large scale magnetic 
fields\cite{R11,R12}. For the superconducting string, there is a current along 
the symmtery axis and consequently a magnetic field in the transverse direction.
The geometry is no longer boost invariant. Moss and Poletti \cite{R13} first
investigated the gravitational field of such a string assuming that at large 
distances from the core of the string, the energy stress tensor is dominated by 
the magnetic field alone. Later, Demiansky \cite{R14} investigated the particle
motion in the spacetime given by Moss and Poletti.

Scalar tensor theories, especially the Brans-Dicke(BD) theory of gravity 
\cite{R15}, which is compatible with the Mach's principle,
have been considerably revived in the recent years. It was shown by
La and Steinhardt \cite{R16} that because of the interaction  of the
BD scalar field and the higgs type sector, which undergoes a strongly
first -order phase transition, the exponential inflation as in the
Guth model \cite{R17} could be slowed down to power law one. The
"graceful exit" in inflation is thus resolved as the phase transition
completes via bubble nucleation.

On the other hand, it seems likely, that in the high energy scales,
gravity is not governed by the Einstein's action, and is modified by
the superstring terms which are scalar tensor in nature. In the low
energy limit of this string theory, one recovers Einstein's gravity
along with a scalar dilaton field which is non minimally coupled to
the gravity \cite{R18}.Although dilaton gravity and BD theory arise
from entirely different motivations, it can be shown by a simple
transformation of the scalar field that the former is a special case of
the latter at least formally.

The dilaton gravity is given by the action

$$
A = {1\over{16\pi}}\int \sqrt{-g} e^{-2\phi}(R +
4g^{\mu\nu}\phi_{,\mu}\phi_{,\nu}) d^{4}X,
\eqno{(1a)}
$$

where no matter field is present except the massless dilaton field $\phi$.

Now if one defines a variable

$$
\psi = e^{-2\phi},
\eqno{(1b)}
$$

then the action (1a) looks like

$$
A = {1\over{16\pi}} \int \sqrt{-g}(\psi R +
{\psi_{,\mu}\psi_{,\nu}\over{\psi}} g^{\mu\nu}) d^{4}X,
\eqno{(1c)}
$$

which is indeed a special case of BD theory given by the action

$$
A = {1\over{16\pi}} \int \sqrt{-g}(\psi R -
\omega{\psi_{,\mu}\psi_{,\nu}\over{\psi}} g^{\mu\nu}) d^{4}X,
\eqno{(1d)}
$$

for the parameter $\omega = -1$.

But when matter field is present, then the action in the dilaton gravity
is given by
$$
A = {1\over{16\pi}}\int \sqrt{-g} e^{-2\phi}(R +
4g^{\mu\nu}\phi_{,\mu}\phi_{,\nu} + e^{2a\phi}{\cal{L}}) d^{4}X,
\eqno{(1e)}
$$
where $\cal{L}$ is the lagrangian for the matter field present. 

The action (1e) cannot be reduced to the corrsponding action for the BD
theory which is given by

$$
A = {1\over{16\pi}} \int \sqrt{-g}(\psi R -
\omega{\psi_{,\mu}\psi_{,\nu}\over{\psi}}g^{\mu\nu} + {\cal{L}}) d^{4}X,
\eqno{(1f)}
$$

by the transformation (1b) and putting $\omega = -1$ because
 the non minimal coupling between the matter lagrangian $\cal{L}$
and the dilaton field $\phi$ in the action (1e) which is not present in
(1f) for the BD theory.

But for $a = 1$ in (1e), there is no coupling between the $\cal{L}$ and
the dilaton field $\phi$ and one can reduce (1e) to the corresponding BD
action (1f).

The renewed interest in BD theory acquires more points of interest when we
observe that the topological defects such as domain walls, cosmic strings,
and monopoles are formed during inflation in the early phase of the
universe and hence can interact  with the BD scalar field with remarkable
change in their properties \cite{R19,R20}. The implications of the dilaton
gravity for the defects have also been explored by several authors
\cite{R21,R22}.

In the present work, we have investigated the gravitational field of a 
superconducting string in Brans-Dicke (B-D) theory of gravity. Like Moss and
Poletti, we have assumed that in the far field zone, the energy momentum tensor 
for the string is dominated by the magnetic field. We have found a family of 
solutions for the spacetime in the far field zone depending upon the value
of some arbitary constants. One of our solutions reduces to the solution 
given by Moss and Poletti \cite{R13} when the B-D scalar field becomes constant,
but for other case one cannot recover the corresponding GR solution.

The gravitational field equations in B-D theory are given by
$$
G_{\mu\nu}=8\pi{T_{\mu\nu}\over{\phi}}+{\omega\over{\phi^{2}}}(\phi_{,\mu}\phi_{,\nu}-{1\over{2}}g_{\mu\nu}\phi_{,\alpha}\phi^{,\alpha})+
{1\over{\phi}}(\phi_{,\mu;\nu}-g_{\mu\nu}\Box\phi),
\eqno{(2a)}
$$
and\\
$$
\Box\phi={8\pi T\over{(2\omega+3)}}
\eqno{(2b)}
$$
in units where $c=1$. $T_{\mu\nu}$ is the energy momentum tensor representing
the contribution from any other field except the B-D scalar field $\phi$ and
$\omega$ is the constant parameter. $T$ represents the trace of $T_{\mu\nu}$.

To describe the spacetime geometry due to an infinitely long static cosmic 
string, the line element is taken to be general static cylindrically symmetric
one given by,
$$
ds^{2}= e^{2(K-U)}(-dt^{2}+dr^{2})+e^{2U}dz^{2}+e^{-2U}W^{2}d\theta^{2},
\eqno{(3)}
$$
where $K,U,W$ are functions of the radial coordinate $r$.

A model for the fields of superconducting cosmic string can be described by a
$U(1)\times U(\tilde{1})$ gauge theory with the gauge fields $A_{\mu}$ and $B_{\mu}$
are coupled to the scalar fields $\sigma$ and $\psi$ \cite{R9,R13}. The string 
is represented by the Nielsen-Olesen \cite{R23} vortex solution with
$$
\psi= \psi(r)e^{i\theta}, \hspace{2mm} B_{\mu}=(1/e)[B(r)-1]\delta^{\theta}_{\mu}
\eqno{(4a)}
$$
where $e$ is the coupling parameter.

Outside the core of the string, the $U(\tilde{1})$ symmetry is broken and 
$\psi$ attains a nonzero expectation value at the minimum of the potential
$V(\sigma,\psi)$.

The $U(1)$ symmetry is unbroken away from the string with $A_{\mu}$ representing 
photon. We have choosen \cite{R13},
$$
\sigma=\sigma_{0}(r)e^{i\chi}, \hspace{2mm} A_{\mu}=A(r)\delta^{z}_{\mu}
\eqno{(4b)}
$$
with the phase $\chi$ being a function of $z$.

With (4b), there is an electric current density $J$ along the z-axis, given by
$$
J= 2ej\sigma^{2}e^{-2U}
\eqno{(5)}
$$
where
$$
j={\partial\chi\over{\partial z}}+ eA
\eqno{(6)}
$$
The Maxwell equation which we will need later is 
$$
({\sqrt{-g}}e^{-2K}A^{\prime})^{\prime} = 8\pi ej\sigma_{0}^{2}e^{-2U}{\sqrt{-g}}
$$
Which on integration yields a first integral
$$
A^{\prime} = {2Ie^{2U}\over{W}}
\eqno{(7)}
$$
$$
with \hspace{2mm} I(r)=\int{J 2\pi \sqrt{-g} dr}
\eqno{(8)}
$$
being the electric current along the symmetry axis of the string.

As one moves away from the string, the all other string fields drop off 
rapidly  leaving the magnetic field $A_{\mu}$ of its own. The energy momentum
tensor then takes the form
$$
T^{t}_{t}=-T^{r}_{r}=-T^{z}_{z}=T^{\theta}_{\theta}= {1\over{2}}A^{\prime 2}e^{-2K}
\eqno{(9)}.
$$
With (9) and (3), the B-D field equations (2) take the form
$$
(W^{\prime}\phi)^{\prime}=(\phi^{\prime}W)^{\prime}=0
\eqno{(10a)}
$$
$$
(K^{\prime}W\phi)^{\prime}=0
\eqno{(10b)}
$$
$$
2U^{\prime\prime}+{2U^{\prime}W^{\prime}\over{W}}-2K^{\prime\prime}-2U^{\prime 2}=
\omega{\phi^{\prime 2}\over{\phi^{2}}}+2{W^{\prime\prime}\over{W}}
\eqno{(10c)}
$$
$$
(U^{\prime}W\phi)^{\prime}=-{4\pi A^{\prime 2}e^{-2U}W}
\eqno{(10d)}
$$
Here prime denotes differentiation w.r.t $r$.

Solving (10a) and (10b) we get
$$
\phi = \phi_{0}r^{p}
\eqno{(11a)}
$$
$$
W = W_{0}r^{1\over{n}}
\eqno{(11b)}
$$
and
$$
e^{K} = \alpha r^{b\over{n}}
\eqno{(11c)}
$$
where $\phi_{0}, W_{0}, b, \alpha,$ and $n$ are constants of integration and
$p={{n-1}\over{n}}$.

Using (11a), (11b) and (11c), one gets from (10c)
$$
U^{\prime\prime} = U^{\prime 2}-{U^{\prime}\over{nr}}+{M\over{r^{2}}}
\eqno{(12)}
$$
where
$$
M={1\over{2}}[\omega p^{2} +{2(1-n)\over{n^{2}}}-{2b\over{n}}]
\eqno{(12a)}
$$

One can have three solutions of (12) depending on $(1-4C)$ is +ve, zero, or -ve,
where $C=M+{1\over{4n^{2}}}$. In the follwing Calculations,
we have concentrated in the casess where $(1-4C)$ is +ve and zero.

Case 1 {\underline{$1-4C>0$}}

In this case the solution of (12) becomes
$$
U = -ln[C_{1}r^{i_{1}}+C_{2}r^{i_{2}}]
\eqno{(13)}
$$
where $C_{1}$ and $C_{2}$ are arbitary constants and 
$$
i_{1} = {1\over{2}}+{\sqrt{1-4C}\over{2}}-{1\over{2n}}
\eqno{(14a)}
$$
$$
i_{2} = {1\over{2}}-{\sqrt{1-4C}\over{2}}-{1\over{2n}} 
\eqno{(14b)}
$$

Hence the complete line element in the far field zone of a superconducting 
string in B-D theory becomes
$$
ds^{2} = \alpha^{2}r^{2b\over{n}}H^{2}(-dt^{2}+dr^{2})+H^{-2}dz^{2}+
W_{0}^{2}r^{2\over{n}}H^{2}d\theta^{2}
\eqno{(15)}
$$
where $H=C_{1}r^{i_{1}}+C_{2}r^{i_{2}}$.

Putting (11a), (11b), (11c) and (13) in eqn(10d) and using (7) one can get the 
relation
$$
C_{1}C_{2}(i_{1}-i_{2})^{2}\phi_{0}W_{0}^{2}=16\pi I^{2}
\eqno{(16)}
$$
Now if one puts $n=1$ and $\phi_{0}={1\over{G}}$ where $G$ is the gravitational 
constant, it is easy to verify that the B-D scalar field becomes constant and
$$
ds^{2} = \alpha^{2}r^{2b}F^{2}(-dt^{2}+dr^{2})+F^{-2}dz^{2}+
W_{0}^{2}r^{2}F^{2}d\theta^{2}
\eqno{(17)}
$$
where $F=C_{1}r^{\sqrt{b}}+C_{2}r^{-\sqrt{b}}$
and (16) becomes
$$
C_{1}C_{2}={4\pi GI^{2}\over{bW_{0}^{2}}}
\eqno{(18)}
$$
Equations (17) and (18) are the result earlier obtained by Moss and Poletti
\cite{R13} in GR.Comparing our result to the result earlier obtained by Moss 
and Ploetti one can identify the two integration constants $b,n$  to be related
to the Mass per unit length of the string and the field energy density of the 
string respectively.

One can also calculate in the spacetime given by (15), the radial accelaration
$\dot{v}^{1}$ of a particle that remains stationary (i.e. $v^{1}=v^{2}=v^{3}=
0$) in the field of the string.

Now
$$
\dot{v}^{1} = v^{1}_{;0}v^{0} = \Gamma^{1}_{00}v^{0}v^{0}
\eqno{(19a)}
$$
In our spacetime (15) 
$$
v^{0}={1\over{\alpha r^{b\over{n}}[C_{1}r^{i_{1}}+C_{2}r^{i_{2}}]}}
\eqno{(19b)}.
$$
Using (19b), we get from (19a)
$$
\dot{v}^{1}={b(C_{1}r^{i_{1}}+C_{2}r^{i_{2}})+
n(C_{1}i_{1}r^{i_{1}}+C_{2}i_{2}r^{i_{2}})\over{\alpha^{2}nr^{(2b/n+1)}
[C_{1}r^{i_{1}}+C_{2}r^{i_{2}}]^{3}}}
\eqno{(20)}
$$
Now if one assumes $b>0$ and $n>0$, then $\dot{v}^{1}>0$. So the particle
has to accelarate away from the string, which implies that the gravitational
force due to the string itself is attractive. It is similar to the case of 
superconducting string in GR \cite{R13}. But for other values of $b,n$, the string 
may have a repulsive effect.

\vspace{5mm}
Case 2 {\underline{1-4C=0}}

In this case the solution of (12) becomes
$$
e^{-U} = r^{{1\over{2}}(1-{1\over{n}})}[C_{1}+C_{2}lnr]
\eqno{(21)}
$$
Hence the line element becomes
$$
ds^{2} = \alpha^{2}r^{2b\over{n}}r^{(1-{1\over{n}})}P^{2}(-dt^{2}+dr^{2})+
r^{-(1-{1\over{n}})}P^{-2}dz^{2}+W_{0}^{2}r^{(1+{1\over{n}})}P^{2}d\theta^{2}
\eqno{(22)}
$$
with $P=C_{1}+C_{2}lnr$ and $4bn=2\omega(n^{2}-2n+1)-4n-n^{2}+5$.

In this case one can get after some straightforward calculations, the result
$$
16\pi I^{2}=-C_{2}^{2}W_{0}^{2}\phi_{0}n
\eqno{(23)}
$$
Now as $\phi_{0}$ should be +ve to ensure the positivity of $G$, the gravitational
constant, $n$ must be -ve. But one can check from the equation (11a) that in
order to make the B-D scalar field $\phi$ a constant i.e. to get the corresponding
GR solution one should put $n=1$. Hence in this case one cannot recover the 
corresponding GR solution.

In conclusion, this work extends the earlier work by Moss and Poletti regarding 
the gravitational field of a superconducting cosmic string to  the 
Brans-Dicke theory of gravity. The main feature of our solutions in B-D theory
is that one can have a family of solutions for the spacetime of the 
superconducting string depending on the choices of arbitary constants.

The author is grateful to Dr. Narayan Banerjee for valuable suggestions. The
author is also grateful to University Grants Commission, India for the financial
support.


\newpage
\begin{thebibliography}{50}
\bibitem{R1}T.W.B.Kibble Phys.Rep {\bf{67}} (1980) 183.
\bibitem{R2}A.Vilenkin Phys.Rev.D {\bf{23}} (1981) 852.
\bibitem{R3}W.A.Hiscock Phys.Rev.D {\bf{31}} (1985) 3288.
\bibitem{R4}G.R.Gott Astrophys.J {\bf{288}} (1985) 422.
\bibitem{R5}D.Harari and P.Sikivie Phys.Rev.D {\bf{37}} (1988) 3448.
\bibitem{R6}R.Gregory Phys.Lett.B {\bf{215}} (1988) 663.
\bibitem{R7}A.D.Cohen and D.B.Kaplan Phys.Lett.B {\bf{215}} (1988) 65.
\bibitem{R8}A.Banerjee, N.Banerjee and A.A.Sen Phys.Rev.D {\bf{53}} (1996) 5508.
\bibitem{R9}E.Witten Nucl.Phys.B {\bf{249}} (1985) 557.
\bibitem{R10}P.Peter Phys.Rev.D {\bf{49}} (1994) 5052.
\bibitem{R11}T.Vachaspati Phys.Lett.B {\bf{265}} (1991) 258.
\bibitem{R12}R.H.Bradenberger, A.C.Davis, A.M.Matheson and M.Trodden Phys.Lett.B
 {\bf{293}} (1992) 287.
\bibitem{R13}I.Moss and S.Poletti Phys.Lett.B {\bf{199}} (1987) 34.
\bibitem{R14}M.Demiansky Phys.Rev.D {\bf{38}} (1988) 698.
\bibitem{R15}C.Brans and R.H.Dicke Phys.Rev. {\bf{124}} (1961) 925.
\bibitem{R16}D.La and P.J.Steinhardt Phys.Rev.Lett.  {\bf{62}} (1989) 376.
\bibitem{R17}A.H.Guth Phys.Rev.D {\bf{23}} (1987) 347.
\bibitem{R18}M.B.Green, J.H.Schwarz and E.Witten {\bf{Superstring Theory}},
(Cambridge University Press, Cambridge, 1987).
\bibitem{R19}C.Gundlach and M.Ortiz Phys.Rev.D {\bf{42}} (1990) 2521;\\
A.Barros and C.Romero J.Math.Phys {\bf{36}} (1995) 5800;\\
A.A.Sen N.Banerjee and A.Banerjee Phys.Rev.D {\bf{56}} (1997) 3706.
\bibitem{R20}A.Barros and C.Romero gr-qc/9707040;\\
A.Banerjee, S.Chatterjee, A.Beesham and A.A.Sen, Class.Quantum.Grav.
{\bf{15}} (1998) 645.
\bibitem{R21}T.Damour and A.Vilenkin Phus.Rev.Lett. {\bf{78}} (1997) 2288;\\
R.Gregory and C.Santos Phys.Rev.D {\bf{56}} (1997) 1194;\\
O.Dando and R.Gregory gr-qc/9802015.
\bibitem{R22}O.Dando and R.Gregory gr-qc/9709029.
\bibitem{R23}H.B.Nielsen and P.Olesen Nucl.Phys.B {\bf{61}} (1973) 45.
\end{thebibliography}
\end{document}